\documentstyle{mn}

\newif\ifAMStwofonts

\ifoldfss
  \ifCUPmtlplainloaded \else
    \NewTextAlphabet{textbfit} {cmbxti10} {}
    \NewTextAlphabet{textbfss} {cmssbx10} {}
    \NewMathAlphabet{mathbfit} {cmbxti10} {} 
    \NewMathAlphabet{mathbfss} {cmssbx10} {} 
  \fi
  \ifAMStwofonts
    \ifCUPmtlplainloaded \else
      \NewSymbolFont{upmath} {eurm10}
      \NewSymbolFont{AMSa} {msam10}
      \NewMathSymbol{\upi}     {0}{upmath}{19}
      \NewMathSymbol{\umu}     {0}{upmath}{16}
      \NewMathSymbol{\upartial}{0}{upmath}{40}
      \NewMathSymbol{\leqslant}{3}{AMSa}{36}
      \NewMathSymbol{\geqslant}{3}{AMSa}{3E}

    \fi
  \fi
\fi 

\ifnfssone
  \newmathalphabet{\mathit}
  \addtoversion{normal}{\mathit}{cmr}{m}{it}
  \addtoversion{bold}{\mathit}{cmr}{bx}{it}
  \newmathalphabet{\mathbfit} 
  \addtoversion{normal}{\mathbfit}{cmr}{bx}{it}
  \addtoversion{bold}{\mathbfit}{cmr}{bx}{it}
  \newmathalphabet{\mathbfss} 
  \addtoversion{normal}{\mathbfss}{cmss}{bx}{n}
  \addtoversion{bold}{\mathbfss}{cmss}{bx}{n}
  \ifAMStwofonts
    \ifCUPmtlplainloaded \else
      \UseAMStwoboldmath
      \makeatletter
      \new@mathgroup\upmath@group
      \define@mathgroup\mv@normal\upmath@group{eur}{m}{n}
      \define@mathgroup\mv@bold\upmath@group{eur}{b}{n}
      \edef\UPM{\hexnumber\upmath@group}
      \new@mathgroup\amsa@group
      \define@mathgroup\mv@normal\amsa@group{msa}{m}{n}
      \define@mathgroup\mv@bold\amsa@group{msa}{m}{n}
      \edef\AMSa{\hexnumber\amsa@group}
      \makeatother
      \mathchardef\upi="0\UPM19
      \mathchardef\umu="0\UPM16
      \mathchardef\upartial="0\UPM40
      \mathchardef\leqslant="3\AMSa36
      \mathchardef\geqslant="3\AMSa3E
    \fi
  \fi
\fi 

\ifnfsstwo
  \DeclareMathAlphabet{\mathbfit}{OT1}{cmr}{bx}{it}
  \SetMathAlphabet\mathbfit{bold}{OT1}{cmr}{bx}{it}
  \DeclareMathAlphabet{\mathbfss}{OT1}{cmss}{bx}{n}
  \SetMathAlphabet\mathbfss{bold}{OT1}{cmss}{bx}{n}
  \ifAMStwofonts
    \ifCUPmtlplainloaded \else
      \DeclareSymbolFont{UPM}{U}{eur}{m}{n}
      \SetSymbolFont{UPM}{bold}{U}{eur}{b}{n}
      \DeclareSymbolFont{AMSa}{U}{msa}{m}{n}
      \DeclareMathSymbol{\upi}{0}{UPM}{"19}
      \DeclareMathSymbol{\umu}{0}{UPM}{"16}
      \DeclareMathSymbol{\upartial}{0}{UPM}{"40}
      \DeclareMathSymbol{\leqslant}{3}{AMSa}{"36}
      \DeclareMathSymbol{\geqslant}{3}{AMSa}{"3E}
    \fi
  \fi
\fi 

\ifCUPmtlplainloaded \else
  \ifAMStwofonts \else 
    \def\upi{\pi}
    \def\umu{\mu}
    \def\upartial{\partial}
  \fi
\fi

\title[The Effect of a Central Supermassive Black Hole]
{The Effect of a Central Supermassive Black Hole on the Gas Fuelling}
\author[H. Fukuda, K. Wada and A. Habe]
{Hiroyuki Fukuda,$^1$ \thanks{fukuda@phys.hokudai.ac.jp} Keiichi Wada$^2$ \thanks{wada@th.nao.ac.jp} and Asao Habe$^1$ \thanks{habe@phys.hokudai.ac.jp}\\
$^1$Division of Physics, Graduate School of Science, Hokkaido University, Kita 10 Nishi 8, Sapporo 060, Japan\\
$^2$National Astronomical Observatory, Mitaka, Tokyo 181, Japan}
\date{}

\begin{document}

\maketitle

\begin{abstract}
When a supermassive black hole exists in the centre of a galaxy, an additional
inner Lindblad resonance (ILR) exists inside the usual ILRs. We study gas
dynamics in a weakly barred potential with a central supermassive black hole by
using 2D numerical simulations, and we investigate the effect of the additional
ILR on fuelling gas into nuclear starburst regions or AGNs. Our numerical
results show that strong trailing spiral shocks are formed at the resonance
region, and the gas in the shock region is rapidly fuelled into a central
region and make a nuclear gas ring. As a result, a large amount of gas is
concentrated in the nuclear region beyond the ILR in a dynamical time scale.
\end{abstract}


\section{Introduction}

Nuclear activities in galaxies, such as nuclear starbursts or AGNs, are supposed
to be induced by gas fuelling into nuclear regions of galaxies. Observational
studies have suggested that non-axisymmetric gravitational potential caused by
the stellar bar or galaxy-galaxy interactions is a convincing mechanism for
triggering the gas fuelling (Phinney 1994). This is because the
non-axisymmetric potential can remove the angular momentum of gas, and a large
amount of gas can fall toward the centre in a dynamical time scale.
However, numerical simulations have shown that the gas tends to form an oval
ring-like structure near the inner Lindblad resonance (ILR) (Elmegreen 1994
and references there in). It is also shown that the bar can not force the
gas to accrete toward the galactic centre beyond the ILR. As a mechanism to overcome the ILR barrier, the double barred structure (Friedli \& Martinet 1993),
or the self-gravity of gas (Wada \& Habe 1992, 1995; Elmegreen 1994) are
proposed.

For an alternative mechanism to fuel the gas to the galactic centre we
investigate the effect of a central supermassive black hole (SBH). Recent
observations suggest that some galaxies have central SBHs (Ford et al. 1994;
Harms et al. 1994; Miyoshi et al. 1995). If the central SBH exists, an
additional ILR (hereafter a nuclear Lindblad resonance (NLR) ) appears inside
of the usual ILRs. Therefore, the stellar and gas dynamics in the resonant
region are affected by the NLR, and the gas fuelling into the inside of the
usual ILRs caused by the NLR is expected. Hasan \& Norman (1990) investigated
the orbits of a star in a barred galaxy with a central SBH and showed that the
stochastic regions appear as the central mass is increased, and the orbits
sustaining the bar are dissolved. Pfenniger \& Norman (1990) showed that the
dissipation is enhanced in the resonance region and that the gas inflow is
boosted inside the ILRs (including the NLR). However, since Pfenniger \& Norman
(1990) used weakly dissipative single particles to represent idealised gas
clouds, the effects of the NLR on the dynamics of the actual interstellar
matter is not still clear.

In this paper, we investigate the dynamics of a non-self-gravitating gas disc
near the NLR region of a weakly barred galaxy with the central SBH by using
the smoothed particle hydrodynamics (SPH) method. Our aim is to clarify a role
of the new resonance caused by the central SBH on triggering the gas fuelling
for the galactic centre.

In section 2, we present models of galaxies with a central SBH and a gas disc,
and a numerical method. Numerical results are presented in section 3. In
section 4, we summarise our results and discuss the mechanism of gas fuelling
by the NLR  and the observations related to our numerical results and the
effects of the self-gravity of gas.

\section{Models}

\subsection{Model galaxy}

The numerical method and models are based on Wada \& Habe (1992). We assume
that the model galaxy is composed of a stellar disc, a bulge, a weak stellar
bar, and a supermassive black hole (SBH) at the centre of the galaxy. The weak
stellar bar is treated as an external fixed potential. We do not consider a
dark halo component, because we are interested in gas dynamics in the inner
region of the galaxy.

Axisymmetric potential of the galaxy is assumed to be the Toomre disc (Toomre
1963) :
\begin{equation}
\Phi_{\rm axi}(R)=-\frac{c^2}{a}\frac{1}{(R^2+a^2)^{1/2}},
\end{equation}
where $a$ is a core radius and $c$ is given as
$c=v_{\rm max}(27/4)^{1/4}a$, and $v_{\rm max}$ is a maximum rotational
velocity in this potential.

We assume the barred potential :
\begin{equation}
\Phi_{\rm bar}(R,\theta)=\epsilon(R)\Phi_{\rm axi}\cos2\theta,
\end{equation}
where $\epsilon(R)$ is given by
\begin{equation}
\epsilon(R)=\epsilon_0\frac{aR^2}{(R^2+a^2)^{3/2}},
\end{equation}
and $\epsilon_0$ is a parameter which represents the strength of the bar to the 
disc component.

We assume the potential of the SBH :
\begin{equation}
\Phi_{\rm BH}(R)=-f_{\rm BH}\frac{c^2}{a}\frac{1}{(R^2+{a_{\rm BH}}^2)^{1/2}},
\end{equation}
where $f_{\rm BH}$ is a mass ratio of the SBH to the galaxy, and $a_{\rm BH}$
is a softening parameter.

Unit mass, length, and time are $3{\times}10^{10}$ ${\rm M_{\odot}}$, 5 kpc,
and $1.75\times10^8$ yr, respectively. We chose a = 0.4, and $c^2/a=48.7$, for
which the rotational period at $R=1.0$ is 1.0, and $\epsilon_0=0.1$, and
$a_{\rm BH}$ = 0.01. Free parameters of this model are $f_{\rm BH}$ and
$\Omega_{\rm bar}$, the pattern speed of the bar. We chose $f_{\rm BH}=0.01,
0.001$ and 0.0, and $\Omega_{\rm bar}=7.0, 3.5$ and 1.0. Typical values for
$f_{\rm BH}$ and $\Omega_{\rm bar}$ are 0.01 and 3.5, respectively. We
summarise these parameters used in our four models in Table 1.

The inner Lindblad resonances (ILRs) exist where
\begin{equation}
\Omega_{\rm bar}=\Omega(R)-\frac{\kappa(R)}{2},
\end{equation}
where $\Omega$ is the circular frequency and $\kappa$ is the epicycle frequency
given by
\begin{equation}
\kappa^2=R\frac{{\rm d}\Omega^2}{{\rm d}R}+4\Omega^2.
\end{equation}
Fig. 1 shows the radial change of $\Omega-\kappa/2$ for $f_{\rm BH}=0.01,
0.001$, and 0.0, and $\Omega_{\rm bar}=7.0, 3.5$, and 1.0 are also shown. For
$f_{\rm BH}=0.0$ (no SBH exists), only the inner ILR exists for $\Omega_{\rm
bar} < 3.5$. For non-zero $f_{\rm BH}$, the NLR exist at inner region
and has larger radius for larger $f_{\rm BH}$. When $f_{\rm BH}=0.01$, the NLR
exists for $\Omega_{\rm bar} > 2.8$, and for $\Omega_{\rm bar} > 1.2$ when
$f_{\rm BH}=0.001$. The radii of the NLR and the inner ILR for our four models
are shown in Table 1.

\subsection{Gas disc model}

In the smoothed particle hydrodynamics (SPH) code (Gingold \& Monaghan 1982), a
gas disc is represented by a large number of quasi-particles with a certain spatial extent. The SPH particles
are randomly distributed within $R=0.3$ in order to represent uniform density
disc at $t=0$. The initial rotational velocity is given in order to balance
centrifugal force caused by the axisymmetric gravitational potential, since the
distorted potential is very weak. The rotational period at $R=0.3$ is 0.3. We
assume that gas is isothermal and that its temperature is $10^4$ K, which
corresponds that the sound speed $C_{\rm s} \sim 10$ km s$^{-1}$. Although the effect of the
self-gravity of the gas is important for the gas dynamics at the inner region
of galaxies, for the first step in investigating the effect of the resonance
caused by the central supermassive black hole, we do not take the self-gravity
of the gas into account.

\subsection{Numerical method}

The equation of motion of the $i$-th SPH particle are
\begin{equation}
\left\{
\begin{array}{l}
\displaystyle{\frac{{\rm d}{\bmath r}_i}{{\rm d}t}}={\bmath v}_i, \\
\\
\displaystyle{\frac{{\rm d}{\bmath v}_i}{{\rm d}t}}=-\nabla \Phi_{\rm axi}-\nabla\Phi_{\rm bar}-\nabla
\Phi_{\rm BH}-\frac{1}{\rho_i}\nabla(P_i+q_i),
\end{array}\right.
\end{equation}
where ${\bmath r}_i$ is the position vector of the $i$-th particle,and
${\bmath v}_i$ is velocity vector of the $i$-th particle, $\rho_i$, $P_i$ and
$q_i$ is density, pressure and artificial viscosity of the gas respectively.

In the SPH method, physical quantities are represented by sum of the smoothing
kernel $W(r,h)$. Therefore, the physical quantity $f$ is represented as,
\begin{equation}
f({\bmath r})=\sum_{j}m\frac{f_j}{\rho_j}W(|{\bmath r}-{\bmath r}_j|,h_j),
\end{equation}
where $m$ is the mass of the SPH particle, and $h_i$ is the particle size. The
$h_i$ is varied depending on the local gas density as $h_i \propto \rho_i^{-1/3}
$. We chose the Gaussian smoothing kernel,
\begin{equation}
W(r,h)=\frac{e^{-r^2/h^2}}{\pi^{3/2}h^3}.
\end{equation}

The pressure gradient is represented as a symmetric form
\begin{equation}
\left(\frac{\nabla P}{\rho}\right)_i=\sum_{j}m\left(\frac{P_i}{\rho_i^2}+\frac{P_j}{\rho_j^2}\right)\nabla_i W(r_{ij},h_{ij}),
\end{equation}
where $h_{ij}=(h_i+h_j)/2$. We chose symmetric form because angular momentum
is not conserved for the non-symmetric form at our test calculation, and is
completely conserved for the symmetric form (see Appendix A).

The artificial viscosity is given by
\begin{eqnarray}
\left(\frac{\nabla q}{\rho}\right)_i=\sum_{j}(-\alpha\mu_{ij}+\beta\mu_{ij}^2)m\nonumber\\
\times\left(\frac{P_i}{\rho_i^2}+\frac{P_j}{\rho_j^2}\right)\nabla_i W(r_{ij},h_{ij}),
\end{eqnarray}
where
\begin{equation}
\mu_{ij}=\frac{2h_{ij}{\bmath v}_{ij}\cdot{\bmath r}_{ij}}{C_{\rm s}(r_{ij}^2+0.1h_{ij}^2)},\mbox{ }\alpha=0.5,\mbox{ }\beta=1.0.
\end{equation}

We use $10^4$ SPH particles. Test calculations of our SPH code are summarised
in Appendix B.

\section{Results}

In this section, we show the time evolution of the gas distribution in our
simulations comparing models with and without a central supermassive black
hole (SBH).

\subsection{A Model without SBH}

In order to emphasize the importance of the nuclear Lindblad resonance (NLR)
caused by the SBH, we first present results of a model without SBH (model N).
The time evolution of the gas disc is as follows (Fig. 2). At first, the gas
disc is distorted in a direction leading the bar major axis by about
$45^\circ$, and two leading spiral shocks are formed at $t=0.3$. Then, the
elongated gas disc begins to rotate clockwise (i.e. retrograde), and becomes
parallel to the bar at $t=0.5$. Since the elongated disc continues to rotate
clockwise, the disc is finally led by the bar. In this phase the gas
disc gain angular momentum form the bar, and finally it becomes almost
axisymmetric. The morphology of the gas disc at $t=1.0$ is almost the same as
that of initial condition. Same as Wada \& Habe (1992), no fuelling toward the
centre occurs in this model. The time evolution of the outer shape of the gas
disc described above is consistent with the adiabatic invariant analysis of a
particle orbit in a barred potential (Lynden-Bell 1979).

Fig. 7 \& 8 show the time evolution of the total angular momentum of gas disc
and of the gas mass within the radius 0.1 for all models. The gas disc of
model N loses angular momentum by the torque from the bar, when the distorted
gas disc leads the bar ($t<0.5$). However, after $t=0.5$ the distorted gas disc
becomes to be led by the bar, that is, the gas receives angular momentum from
the bar. As a result, the total angular momentum recovers the nearly initial
value by the time 1.0. The gas mass within radius 0.1 increases by the effect
of the distortion of the gas disc, then it returns to the almost initial value
when the gas disc returns to be axisymmetric.

\subsection{Models with SBH}

\subsubsection{Model B: $\Omega_{\rm bar}=3.5, f_{\rm BH}=0.01$}

In this model, $f_{\rm BH}=0.01$, which means that the mass of the SBH is about
1 per cent of the mass of a galaxy, and $\Omega_{\rm bar}=3.5$ which is an
upper limit value for the usual ILR to exist (Fig. 1). These value for $f_{\rm
BH}$ and $\Omega_{\rm bar}$ are typical for galaxies. The NLR is located at
$R=0.14$. Fig. 3 shows the time evolution of the gas disc in this model. The
evolution of the gas disc at $t<0.2$ is almost the same as the model without
SBH. However, at $t=0.2$, two trailing spiral shocks which lead the bar about
$45^\circ$ occur, and they extend to the edge of the gas disc. Then, the gas
loses angular momentum substantially at the shocks and it flows drastically
into the centre to make a gas ring with a radius of 0.05. At $t=0.6$, the
elongated gas disc becomes parallel to the bar, then almost all the gas are
fuelled to the central gas ring.

The total angular momentum of gas disc are reduced to about 10 per cent of
the initial value by the time 0.5 (Fig. 7). On the nearly equal time scale,
almost all the gas falls into the central gas ring (Fig. 8).

\subsubsection{Model Ba: $\Omega_{\rm bar}=3.5, f_{\rm BH}=0.001$}

For comparison, we simulate the model with the smaller SBH ($f_{\rm BH}=0.001$)
In this model, the NLR is at $R=0.06$, which is about half of that of model B.
Time evolution of the outer region of the gas disc ($R>0.1$) is almost
the same as model N (Fig. 4). This means that the SBH does not affect the
outer region of the gas disc in this model. Therefore, the total angular
momentum of gas disc evolves quite similar with model N (Fig. 7), because the
outer region of the gas disc is dominant in the total angular momentum.

However, at the inner region ($R<0.1$) of the gas disc, small two trailing
spiral shocks occur, and the gas in the inner region are fuelled into the
centre. The time evolution of the gas mass within $R=0.1$ is similar to
model N (Fig. 8). However, the gas mass within the much inner radius evolves
quite different from model N, and the gas dynamics at the inner region is
almost the same except for model B with the difference of the spatial scale.

\subsubsection{Model Bb: $\Omega_{\rm bar}=7.0, f_{\rm BH}=0.01$}

We calculate the case with the bar rotates so fast that the only NLR
exits (Fig. 5). The NLR is located at $R=0.09$, which is a little smaller than
model B. The outer region of the gas disc ($R>0.2$) evolves like model N,
although the distortion of gas disc is weak, and a time scale of the
pattern rotation of the distorted gas disc is smaller than model N. Fig. 7
shows this property of outer region. This property is because the angular
momentum of outer region of the gas disc is dominant in the total angular
momentum.
The change in the total angular momentum is smaller than model N (Fig. 7), because
the distortion of the gas disc is weaker than model N (Fig. 2 \& 5). Time scale
of the total angular momentum corresponds to the time scale of the pattern
rotation of the distorted disc.

Evolution at the inner region of the gas disc is similar to that in model B
(e.g. trailing spiral shocks almost parallel to the bar). Only the gas
initially located at inner region ($R<0.15$) are fuelled into the central gas
ring. The radius of the gas ring is almost equal to that of model B.
However, the accreted gas mass is small (Fig. 8).

\subsubsection{Model Bc: $\Omega_{\rm bar}=1.0, f_{\rm BH}=0.01$}

We calculate the case with the bar rotates much so slow that no resonance
exits (Fig. 6). In spite of the absence of the resonance, the time evolution of
the gas disc is almost same as model B. Almost all the gas are fuelled into
the much inner region than model B, and time evolution of the total angular
momentum and the gas mass within the radius of 0.1 are also almost the same
with model B (Fig. 7 \& 8). However, trailing spiral shocks in this model leads
the bar about $60^\circ$ which is steeper than model B. This property, the
slower bar induces steeper shocks and faster bar, is consistent of the results
of Wada \& Habe (1995) (see Fig. 12 in Wada \& Habe 1995).

\section{Discussion}

\subsection{Summary of our results}

We have made numerical simulation of the gas discs around a central
supermassive black hole (SBH) in a weak barred potential and have shown the
effects of the nuclear Lindblad resonance (NLR) caused by the SBH. We find that
the gas is highly disturbed by the NLR and trailing spiral shocks are formed
and the gas is finally accumulated into the ring around the SBH whose radius
is about one-third of that of the NLR. The size of the region occupied by the
gas which is accumulated into the gas ring depends on the mass of the SBH or
the pattern speed of the bar. In the models with smaller SBH (model Ba) or
faster bar (model Bb) than model B, only the gas in the smaller region than
model B accretes into the gas ring. In these models, radii of NLR are
also smaller than model B. This suggests that the size of gas region
occupied by the gas which is accumulated is determined by the radius of NLR. It
is notable, however, that the gas fuelling occurs even if the NLR does not
exist (model Bc). Wada (1994) has shown that in such a case, due to the
dissipative nature of gas, highly disturbed motion can be excited. Therefore,
we can explain the strong spiral shocks and, as a result, the rapid gas
fuelling by this process in model Bc.

\subsection{How does the gas fuelling occur ?}

Fig. 9a shows the time evolution of the specific angular momentum of an SPH
particle and a test particle (i.e. pressureless and viscosityless) initially
located near the NLR in model B. Fig. 10 shows the orbits of these
particles. Fig. 9b and 10b are the same as Fig. 9a \& 10a, but for model N.

In model N, the time evolution of the specific angular momentum and the orbit
of the SPH particle are quite similar to that of the test particle (Fig. 9b \&
10b), and the reduction of their angular momentum is small. This is because the
NLR dose not exist. Therefore, their orbits are not so distorted and the
leading spiral shocks in this model is weak. Therefore, angular momentum
and energy reduction of the SPH particle at the shocks is small and the gas
fuelling dose not occur.

In model B, both the SPH particle and the test particle loses specific angular
momentum sufficiently (Fig. 9a), and they take very distorted orbit (Fig. 10a).
Therefore, the trailing spiral shocks in this model are very strong. By the
dissipative nature of the gas, the SPH particle loses angular momentum and
energy at the shocks and then the orbit of the SPH particle bends abruptly at
the shocks. The time and location of the shock passage of the SPH particle are
indicated by arrows in Fig. 9a \& 10a. Then, the SPH particle descends to a
lower angular momentum and lower energy state and, finally, it settles into an
almost circular orbit near the SBH. This means, for whole gas disc, that the
gas fuelling into the nuclear gas ring occurs. The test particle takes chaotic
orbit in this model (Fig. 10a), which is the same result of Hasan \& Norman
(1990).

\subsection{The effect of the sound speed of gas}

Englmaier \& Gerhard (1997) investigated stationary gas flows
in a fixed barred potential, which is similar to our model potential except for
the existence of SBH, and found that the structure of gas flow has two regimes
depending on the sound speed of gas. For the lower sound speed gas, off-axis
shock flow forms and the gas streams inwards through this shock and forms a gas
ring. At the higher sound speed the gas arranges itself in on-axis shock flow
pattern and no gas rings are formed. The critical effective sound speed
dividing the two regimes is around $\sim 20$ km s$^{-1}$ in their standard
model potential and is in the range of values observed in the Milky Way.

Although their results are for the normal ILRs (not NLR), the behaviour of the
gas discs in our models with SBH, which are off-axis shocks and resultant gas
ring, are very similar to that of the lower sound speed regime of them. This is
a reasonable result, because the sound speed of our model is 10 km s$^{-1}$,
and gas responses for the NLR and the OILR are expected to be similar in a
linear analysis (Wada 1994). It should be noticed that the size of the gas
ring formed in our models is much smaller that that in their models due to NLR.

\subsection{Observations}

In our simulations, the nuclear gas ring and the spiral shocks associated with
the ring are formed. Similar gas distribution is observed in nuclear region
of many active galaxies and starburst galaxies (Kenny 1993 for a review).
For example, NGC 4314 has a molecular ring with a radius of about 250 pc and
a star forming region with similar size (Combes et al. 1992). Mini spiral
structures are also observed in this galaxy. These spiral structures are due
to extinction by dust patches (Benedict 1980). In NGC 1097 which is a weak
Seyfert 2 galaxy with barred spiral structure, there is a molecular ring with
the radius of about 700 pc (Gerin, Nakai \& Combes 1988). This galaxy has hot
spots. IC 342 harbours a barlike molecular gas structure $\sim$ 500 pc in extent
(Ishizuki et al. 1990), and the molecular gas structure is very similar to
the trailing spiral shocks in our models. IC 342 also has a modest nuclear
starburst of 70 pc. In our numerical results, in the gas accumulation
phase into the gas ring, violent gas motion is excited and strong shocks occur
in the nuclear region. Since massive gas concentrates in these region, strong
star formation is expected during this phase. Our numerical results can explain
star formation activity in these galaxies.

The trailing spiral shocks formed in our models are short-lived. Therefore,
the observed gas distribution which is similar to that of our models is
expected to be transient structure whose lifetime is about $10^8$yr.

\subsection{Self-gravity of the gas discs}

We estimate the self-gravity of the finally formed gas ring in order to
investigate whether the gas ring is gravitationally unstable or not, since we
do not take into account the self-gravity of the gas in our simulations.
If the gas ring is unstable, the gas ring fragments into clumps and can collide
each other. Then gas fuelling to more inner region will occur (Wada \& Habe
1992, Elmegreen 1994).

We use Toomre's {\it Q} value as the indicator of self-gravitational
instability and take the gas ring formed in model B. Toomre's {\it Q} value of
the gas ring is
\begin{eqnarray}
Q \sim 4\times10^{-3}\frac{\kappa(r_{\rm ring})}{\sigma_{\rm ring}}\left(\frac{C_{\rm s}}{\mbox{10 km s$^{-1}$}}\right) \nonumber \\
 = \frac{0.3}{\sigma_{\rm ring}}\left(\frac{C_{\rm s}}{\mbox{10 km s$^{-1}$}}\right),
\end{eqnarray}
where $\sigma_{\rm ring}$ is the surface density of the gas ring and the radius
of the gas ring $r_{\rm ring}=0.05$. Therefore, if $\sigma_{\rm ring} \ga 0.3$,
$Q \la 1$ and the gas ring is expected to be unstable. Since the surface
density of the gas ring is about 100 times of initial gas disc, the condition
of gravitational instability is $\sigma_{\rm initial} \ga 0.003$ for initial
gas disc. In real units, this condition is that the mass of the initial gas
disc of 1.5 kpc is larger than about $10^7$ ${\rm M_{\odot}}$. Therefore, the
gas ring formed from the gas in a normal galaxy can be gravitationally
unstable, and the self-gravity of the gas ring plays a very important role in
the time evolution of gas ring.

Finally we should note that the self-gravity of the gas at a galaxy centre and
the central SBH play basically the same role on the profile of
$\Omega-\kappa/2$. This is clear from the similarity between Fig. 1 in Wada \&
Habe (1995) and Fig. 1 in this paper. Therefore it is reasonable that the
gaseous evolution and the fuelling process in the both papers are similar.
Self-gravitational effects on a gas disc in a weak bar with a central SBH
should be studied.

\section*{Acknowledgments}

We thank Professor M. Fujimoto for helpful suggestions and a critical
reading of the original version of this paper, and Mr. N. Nakama for fruitful
discussion about orbits around SBH using Lynden-Bell diagram. This work is
supported in part by Research Fellowships of the Japan Society for the
Promotion of Science for Young Scientists (No. 1318), and the Grant-in-Aid for
Scientific Research of the Japanese Ministry of Education, Science, Sports and
Culture (No. 07640344).

{\appendix

\section{The effect of symmetrised formulation}

We test our SPH code from the point of view of angular momentum
conservation. We compare two types of the formulation of the pressure gradient
and the artificial viscosity term. The first one is the non-symmetrised type,
\begin{equation}
\left(\frac{\nabla P}{\rho}\right)_i=\sum_{j}m\frac{P_j}{\rho_i\rho_j}\nabla_i W(r_{ij},h_j),
\end{equation}
which is simply derived from the basic idea of the SPH (Lucy 1977), and the
second one is the symmetric type,
\begin{equation}
\left(\frac{\nabla P}{\rho}\right)_i=\sum_{j}m\left(\frac{P_i}{\rho_i^2}+\frac{P_j}{\rho_j^2}\right)\nabla_i W(r_{ij},h_{ij}),
\end{equation}
We calculate model N by using these two types of pressure-viscosity term
and compare the numerical results. In Fig. 11 , We show the numerical results
of model N with the non-symmetrised code. Until $t \sim 0.3$, distribution of
SPH particles is almost the same with the numerical result of model N with the
symmetrised code. However, after $t \sim 0.7$ evolution of the gas in the
non-symmetrised code is physically incorrect: the dense region becomes unstable
and turns to irregular shape. On the other hand, in the numerical results of
model N with symmetric type the smooth elongated disc is formed in the same
stage. After the stage, the evolution of gaseous structure is completely
different between the non-symmetrised code and the symmetrised code.

In Fig. 12, we show the time evolution of the total angular momentum of the
gas disc of both types. Until $t \sim 0.4$, the time evolution of the total
angular momentum is almost the same in both cases. However, after $t \sim 0.4$
the total angular momentum begins to increase smoothly in the symmetric type
code. On the other hand, total angular momentum in the non-symmetrised code
weakly increase with vibration.

To see the reason of the difference between the codes, we perform the total
angular momentum conservation in a axi-symmetric potential (i.e. without the
bar). In this test, we take initial gas discs from the numerical results of
the non-symmetrised code in the barred potential: gas discs at $t=0.2, 0.3,
0.4,$ and 0.5. Then we calculate the time evolution of the gas discs without
the bar using the symmetrised and non-symmetrised code. The results of the test
are shown in Fig. 11; dotted lines are the results of the non-symmetrised code
and dashed lines are that of the symmetrised code. We found the total angular
momentum conserves completely for all cases with the symmetrised code. However,
the total angular momentum in the non-symmetrised code slightly decrease for
the calculations from the initial gas discs at $t=0.2$ and 0.3, and it is
substantially reduced for the calculations from the initial gas discs at
$t=0.4$ and 0.5. The time $t=0.2$ and 0.3 corresponds to the phase when the
shock regions are small, and the time $t=0.4$ and 0.5 corresponds to the phase
when the shock regions extend to the edge of the disc. Therefore, the
difference of the total angular momentum reduction for the non-symmetrised code
comes from the difference of the extent of shock region. We conclude that the
non-symmetrised  code cannot correctly deal with the shock with high Mach
number. Thus gas dynamics in a rotating weak barred potential is an example
that the symmetrised SPH code must be used.

\section{The effect of numerical shear viscosity}

The artificial viscosity adopted in our SPH code is intend to mimic the bulk
viscosity. However, shear viscosity appears implicitly as pointed by
Hernquist \& Katz (1989). Since the strong shear flows appear in our numerical
results, we have made two critical tests whether the artificial viscosity acts
as a shear viscosity in such a strong shear flow and the angular momentum is
transported artificially. The first test is the comparison of the time
evolution of the gas flows by using our SPH code and a mesh code, and the
second one is the calculation with much larger number of SPH particles.

We calculate model B with AUSM code (Liou \& Steffen 1993; Radespiel \& Kroll
1995), and compared the result with that of SPH code. The special merits of
AUSM compared to the other upwind schemes are low computational complexity and
low numerical diffusion. Fig. 13 shows the velocity fields obtained by using
the both code at $t=0.6$, when the gas fuelling occurs most drastically.
The velocity field of the SPH code in Fig. 13 is calculated on a regular grid,
using the SPH smoothing algorithm. We also show the time evolution of gas mass
within $R=0.1$ for both code in Fig. 14. Both figures show that the two
velocity fields and the time evolution of the gas mass are in good agreement.
Therefore, numerical shear viscosity does not affect the gas dynamics in our
models.

We have also performed the calculation of model B with $5\times10^4$ SPH
particles ( five times larger than our original calculations ). The decrease in
the smoothing lengths will reduce the effects of the artificial transport of
angular momentum. The result of this test is shown in Fig. 14. We compare the
time evolution of gas mass within $R=0.1$ of model B with $10^5$ with the
result with $10^4$ particles in Fig. 14. The difference between two cases is
adequately small. This result supports that the smoothing length in the
calculation with $10^4$ particles is small enough to avoid the artificial shear
viscosity.

Although the mesh code which has less numerical diffusion compared with SPH
code, we have chosen SPH code for this work. This is because high numerical
resolution is naturally achieved in denser region by using SPH, and we have a
plan to develop our study into the gas dynamics involving three dimensional
motion and self-gravity of gas, and the SPH code is easily extended for the
numerical calculation of such a situation.}

\pagebreak

\begin{table}

\caption{Model parameters. $R_{\rm NLR}$ is the radius of the NLR
and $R_{\rm IILR}$ is the radius of the inner ILR.}

\begin{tabular}{ccccc}
\hline
Model & $\Omega_{\rm bar}$ & $f_{\rm BH}$ & $R_{\rm NLR}$ & $R_{\rm IILR}$\\
\hline
N  & 3.5 & 0.0   & not exist & 0.43\\
B  & 3.5 & 0.01  & 0.14 & 0.37\\
Ba & 3.5 & 0.001 & 0.06 & 0.43\\
Bb & 7.0 & 0.01  & 0.09 & not exist\\
Bc & 1.0 & 0.01  & not exist & not exist\\
\hline
\end{tabular}

\end{table}

\begin{figure}

\caption{$\Omega(R)-\kappa(R)/2$ in our models. A thick solid line, a thin
a solid line, and a dashed line are for $f_{\rm BH}=0.01, 0.001$, and 0.0,
respectively. Three horizontal dotted lines represents the pattern speed of
the bars in our models.}

\caption{Time evolution of the distribution of the SPH particles (left) and
the velocity field (right) of gas in model N in the rotating bar
coordinates. The bar major axis is fixed horizontally on each frame.
Time is represented by the number at right up of each figure. A vector
written at right bottom of the figure represents the velocity of 5.0 in out
units. }

\caption{As Fig. 2, but for model B.}

\caption{As Fig. 2, but for model Ba. Only gas distribution is plotted}

\caption{As Fig. 2, but for model Bb. Only gas distribution is plotted.}

\caption{As Fig. 2, but for model Bc. Only gas distribution is plotted.}

\caption{Time variation of the total angular momentum of gas for each model.}

\caption{Time variation of the gas mass in $R<0.1$. Total gas mass in each
model is 1.0.}

\caption{(a) The time evolution of the angular momentum of an SPH particle
initially located near the NLR in model B and a test particle of the same
initial condition. (b) As (a) but for model N. Arrows in Fig. 9a indicates the
time when the SPH particle passes shock region}

\caption{The orbits of the SPH particle and the test particle of Fig. 9 in the
rotating bar coordinates. The bar major axis coincides with the horizontal
axis. (a) for model B. (b) for model N. Arrows in Fig. 10a indicates the
location where the SPH particle passes shock region, and they correspond to the
arrows in Fig. 9a}

\caption{Time evolution of the distribution of the SPH particles in model N
calculated by using the non-symmetrised SPH code.}

\caption{The time evolution of the total angular momentum of the gas disc.
The thick solid line shows the result with the symmetrised code, and the
thin solid line shows the result with the non-symmetrised code.
Dashed lines and dotted lines are the results of the total angular momentum
conservation test. Dashed lines show the results with the symmetrised code, and
the dotted lines show the results with the non-symmetrised code (see Appendix A
for details of the test).}

\caption{The velocity fields at $t=0.6$ obtained by using the AUSM code and SPH
code.}

\caption{Time evolution of the gas mass within $R=0.1$ in model B calculated by
using our SPH code with $10^4$ and $5\times10^4$ SPH particles and AUSM code.
Total gas mass in each model is 1.0.}

\end{figure}

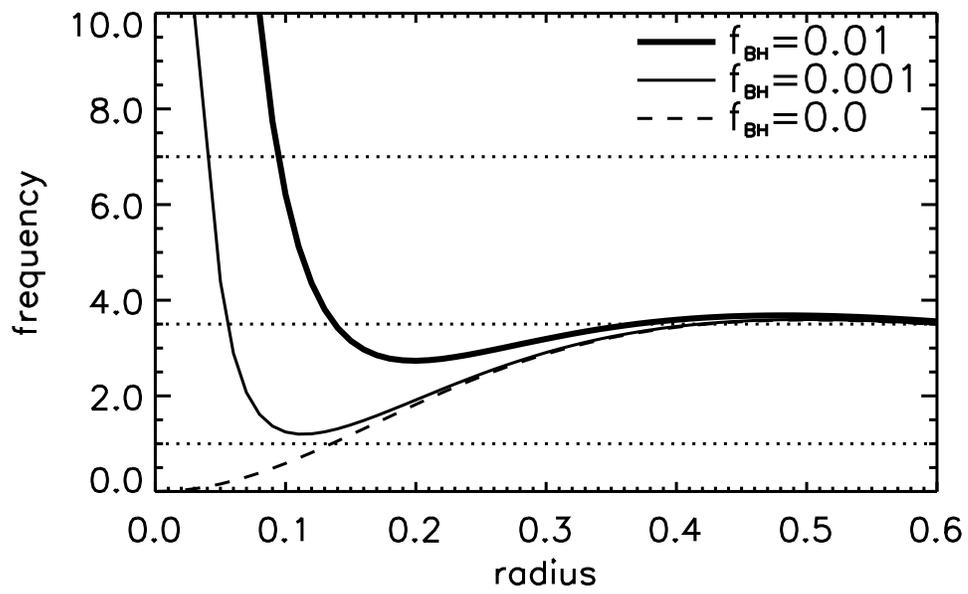
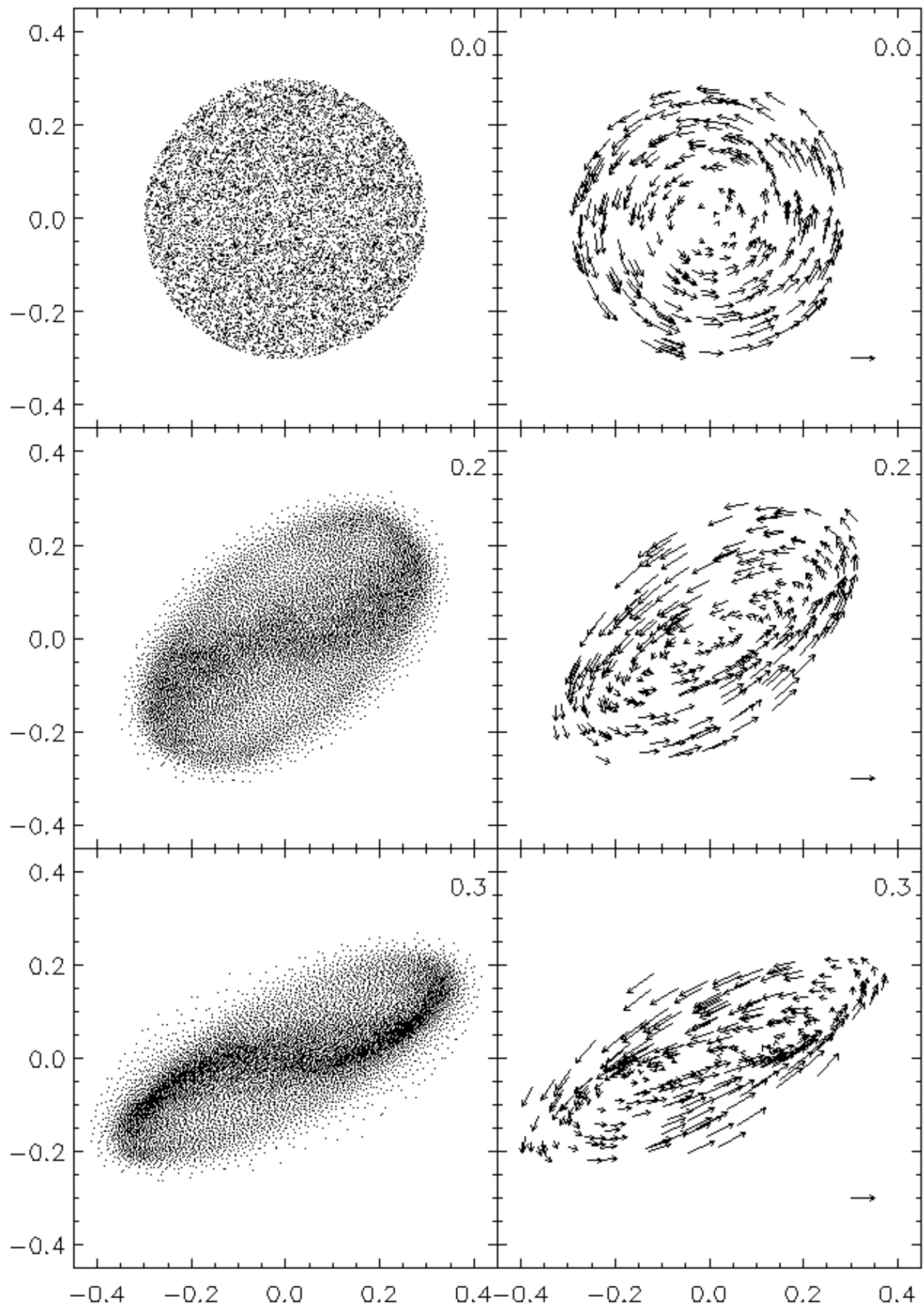
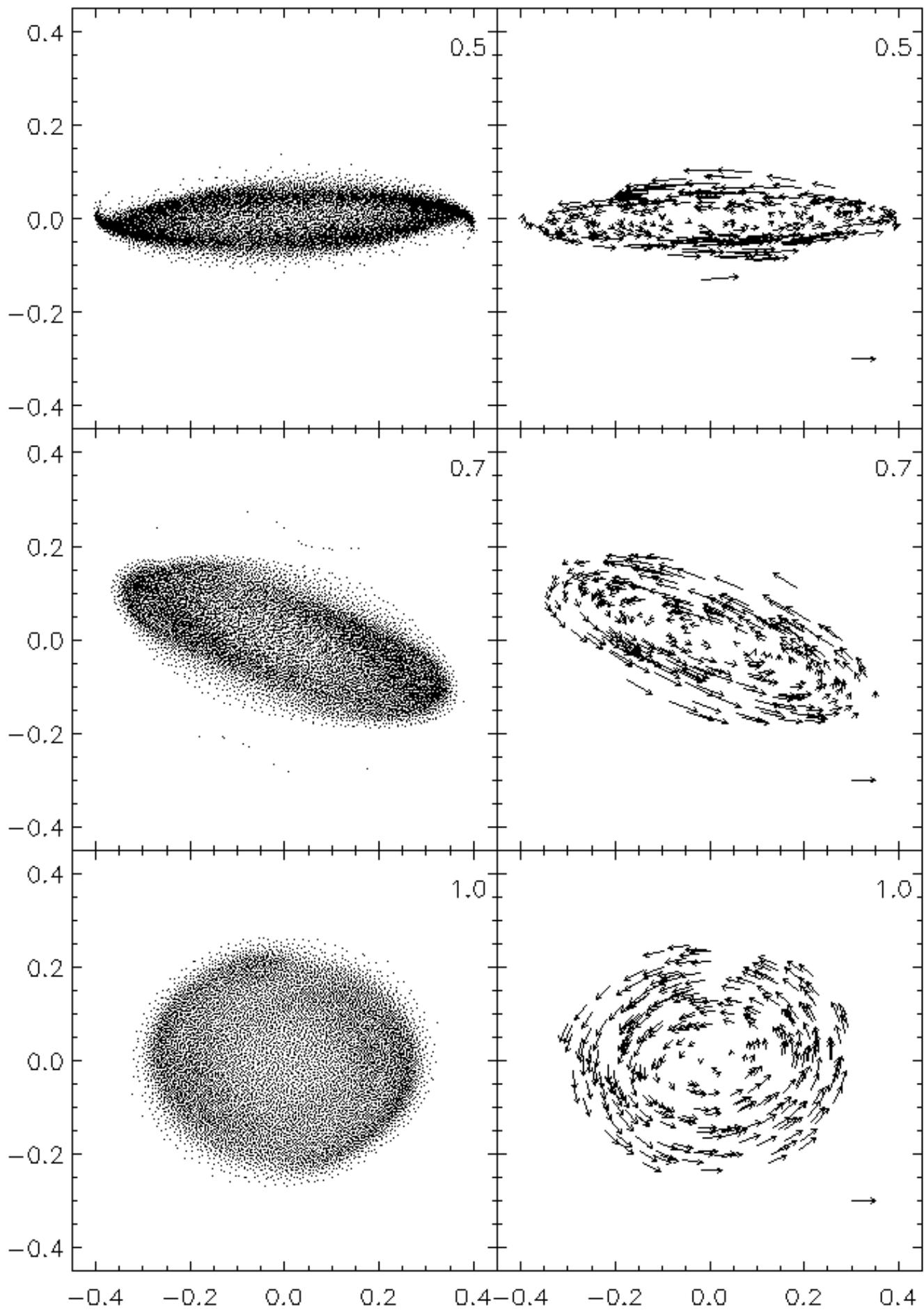
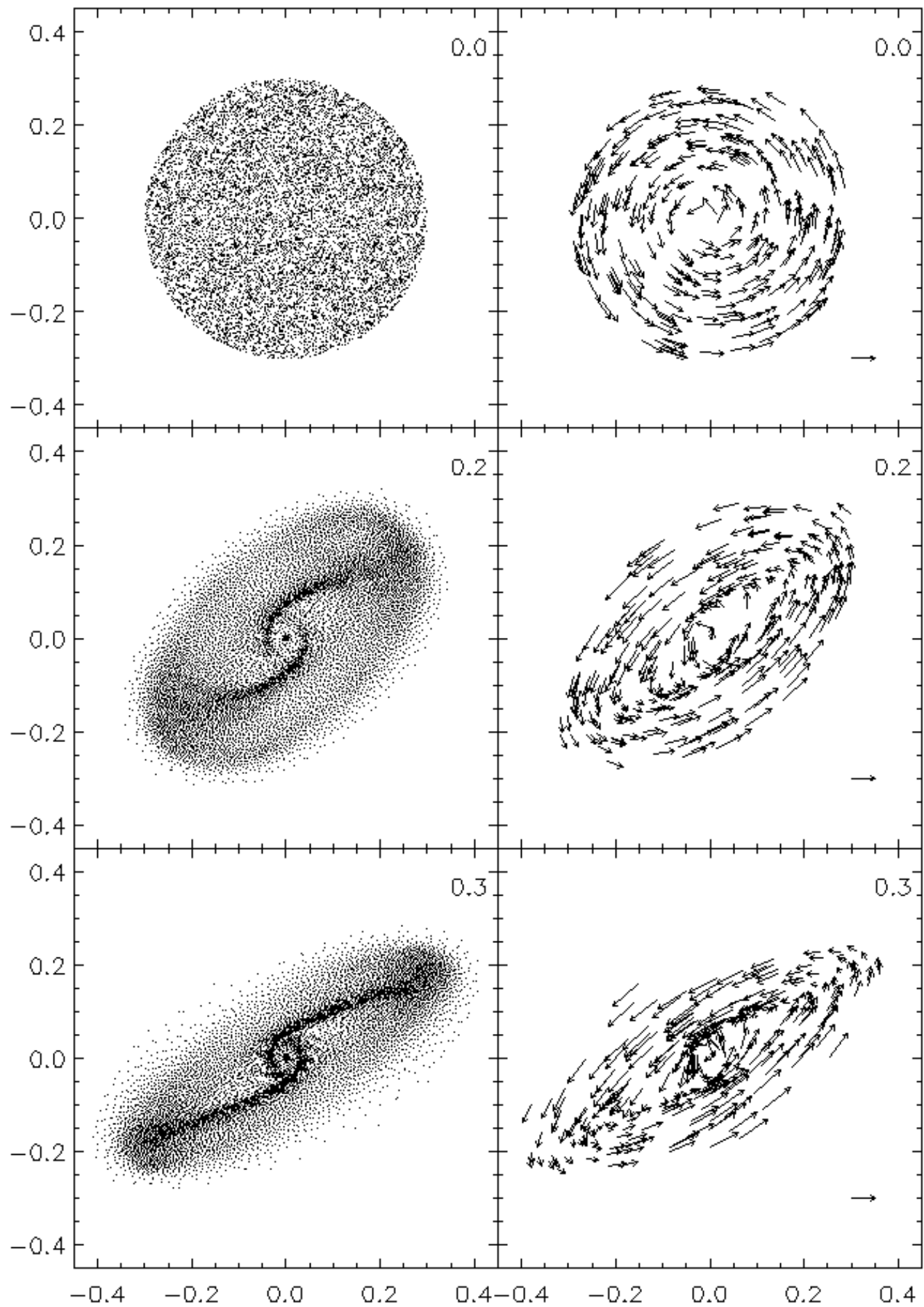
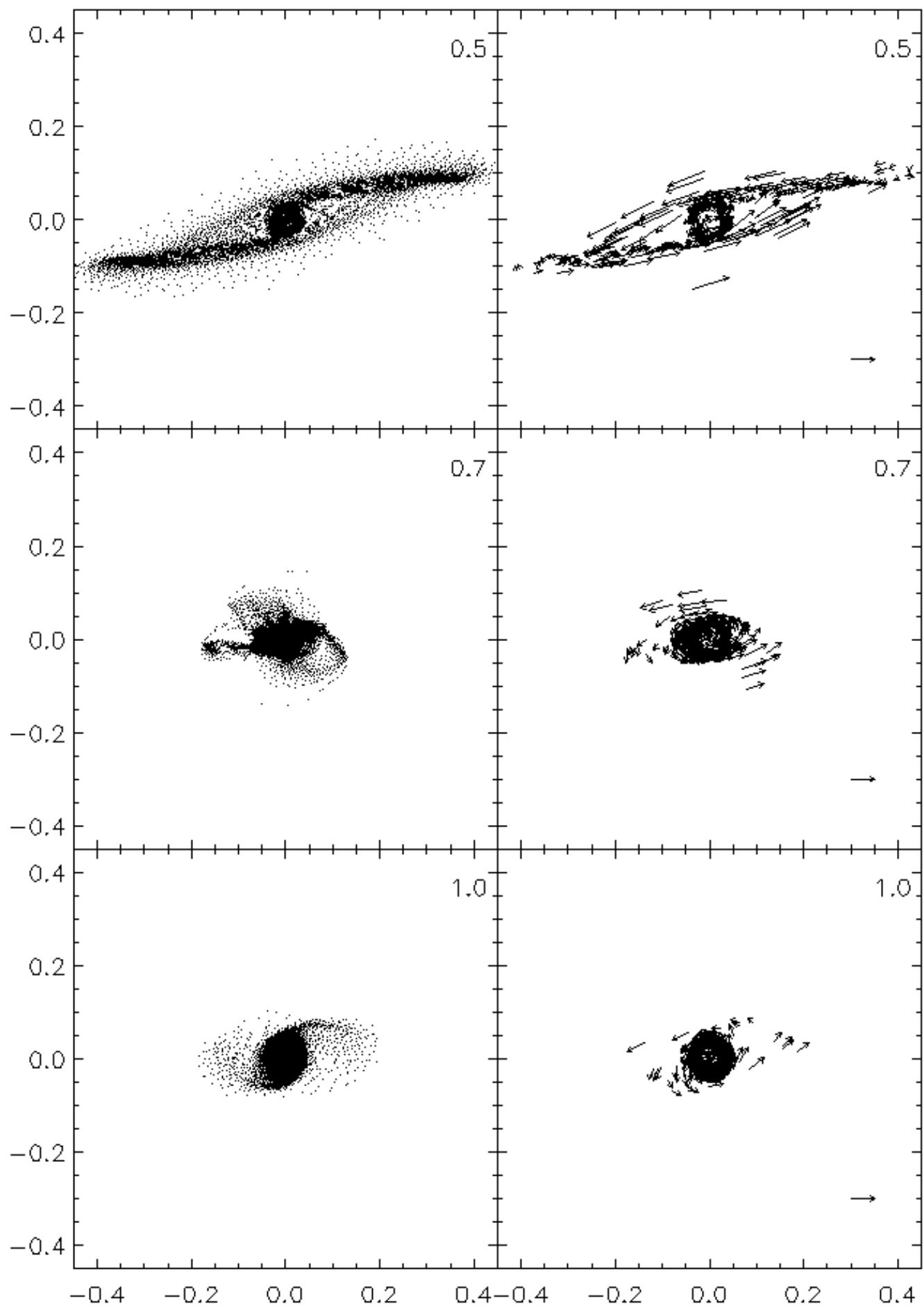
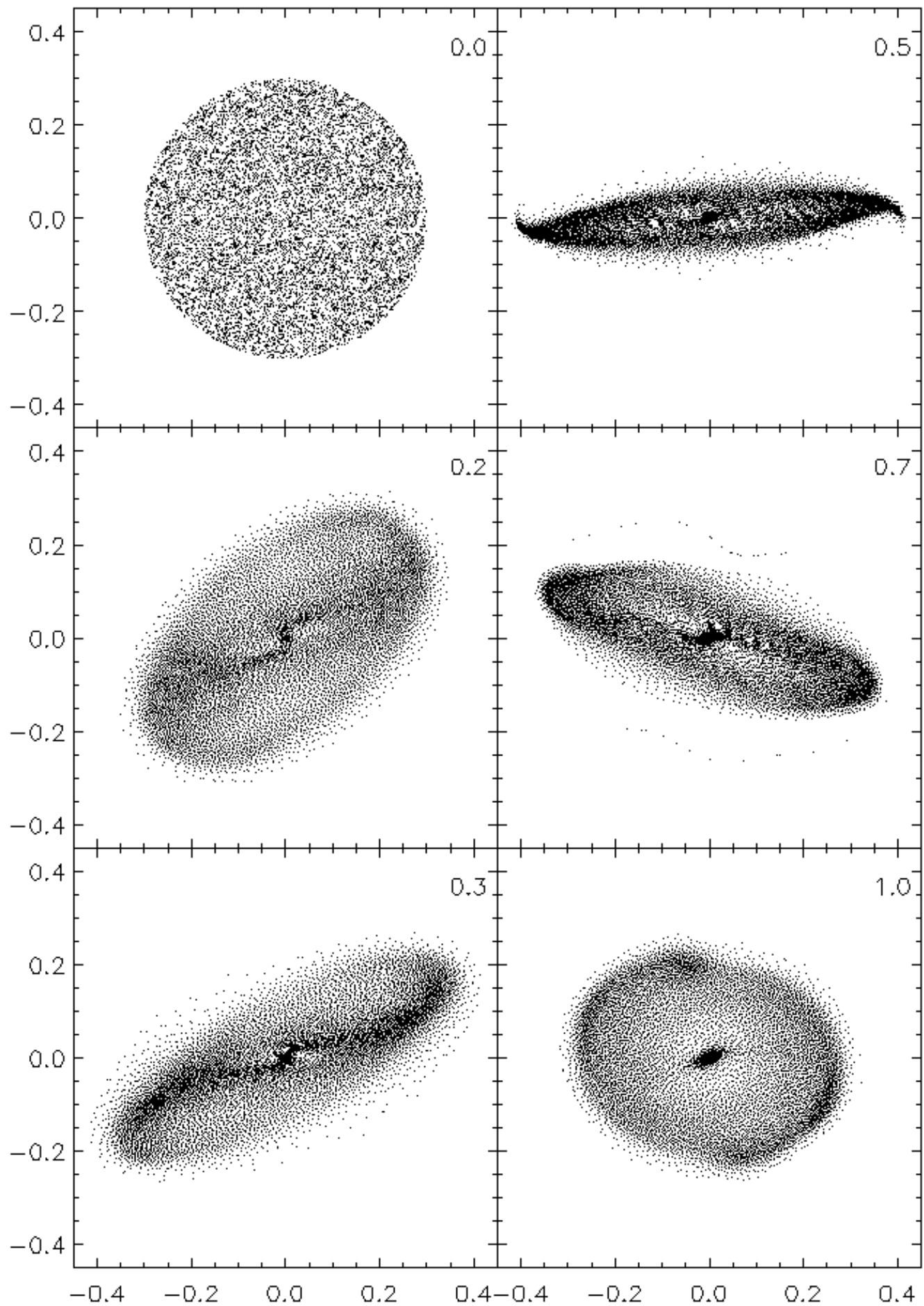
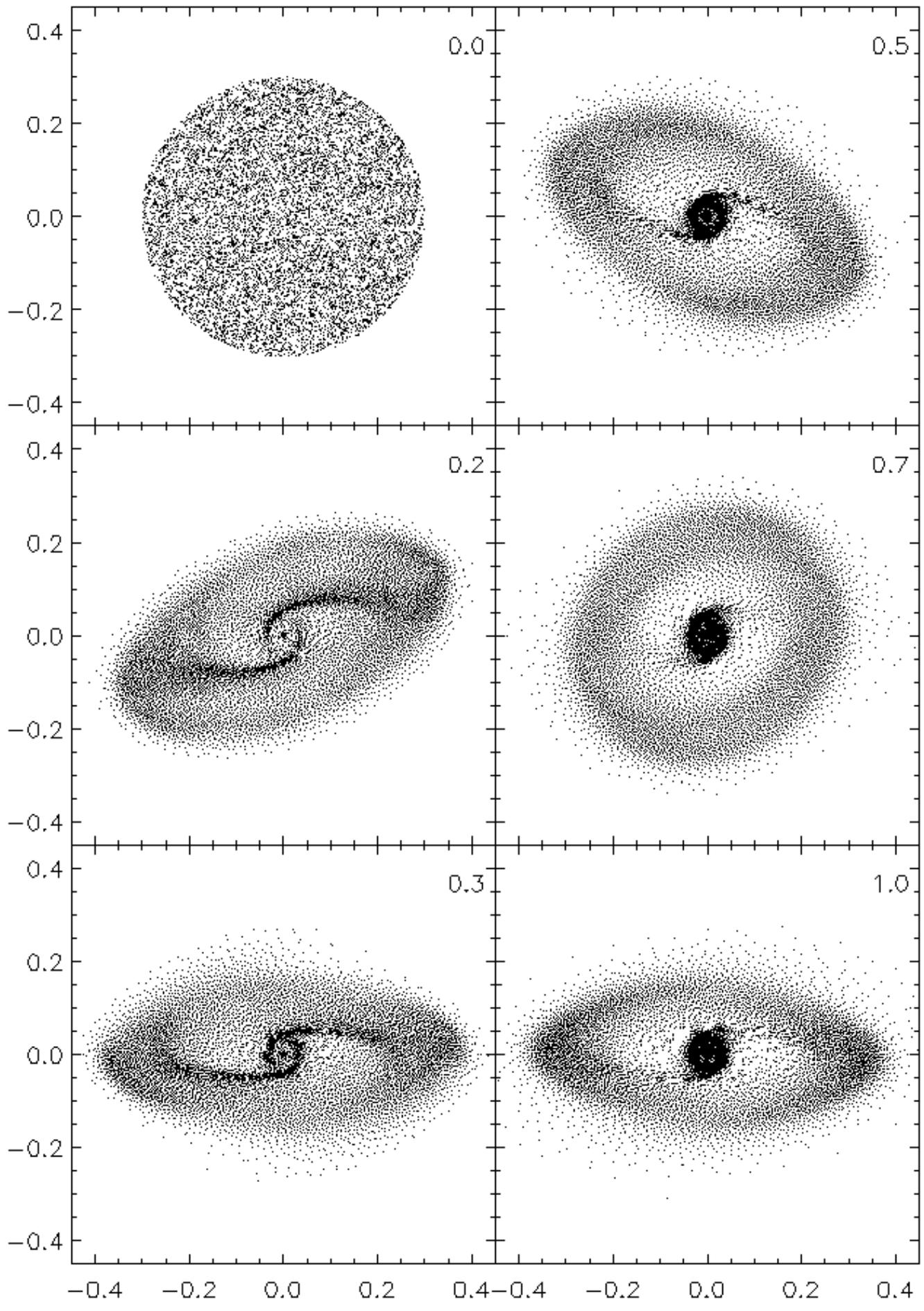
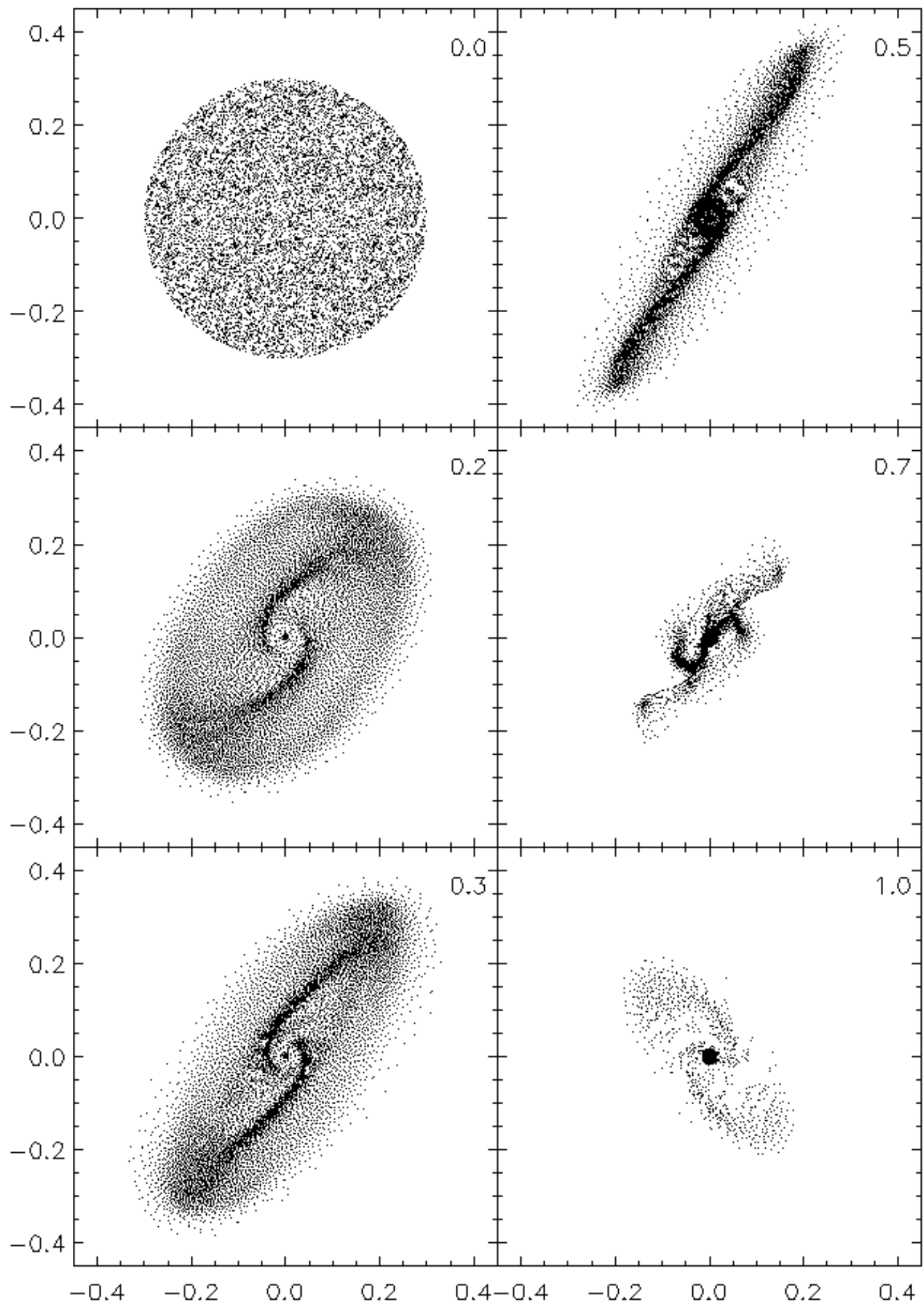
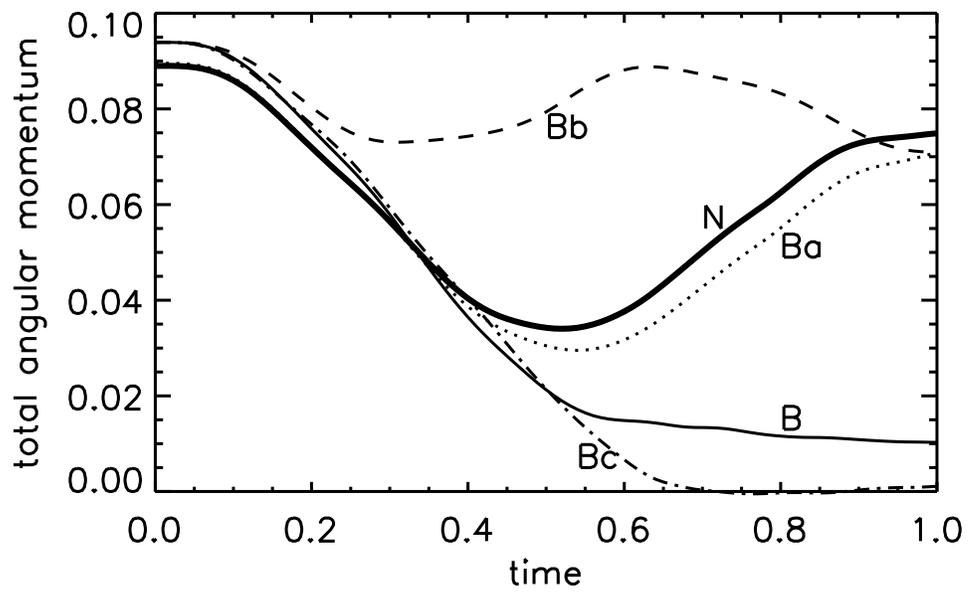
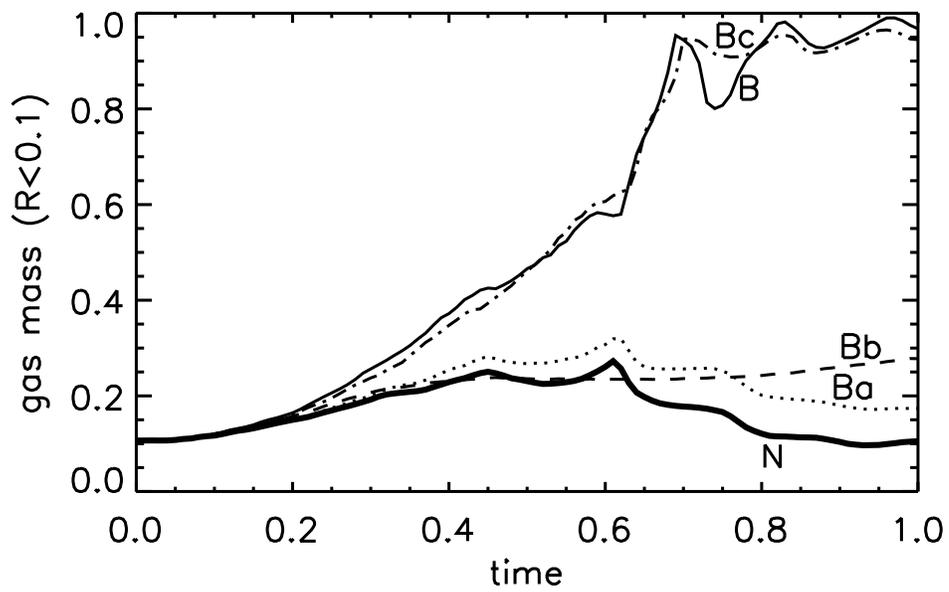
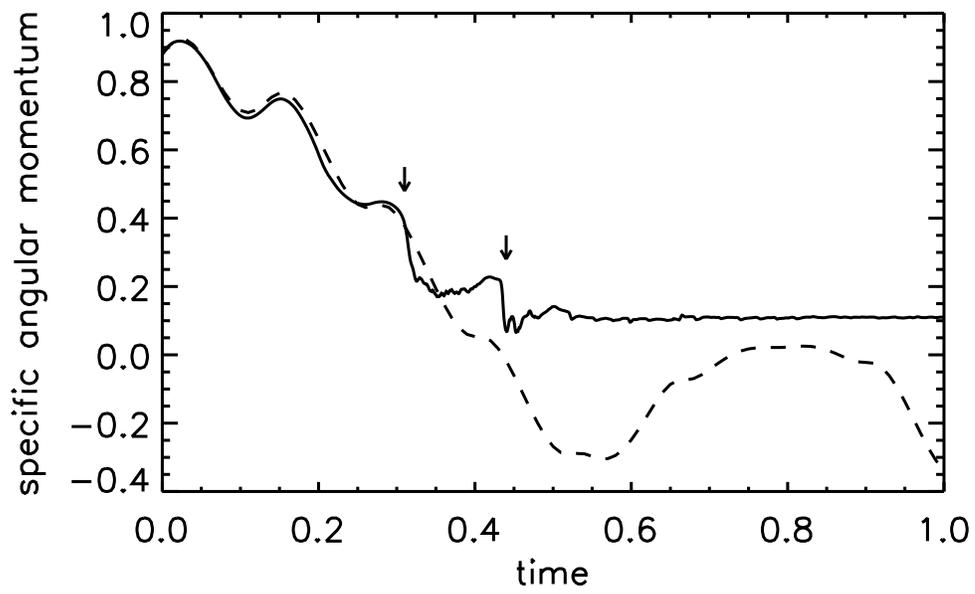
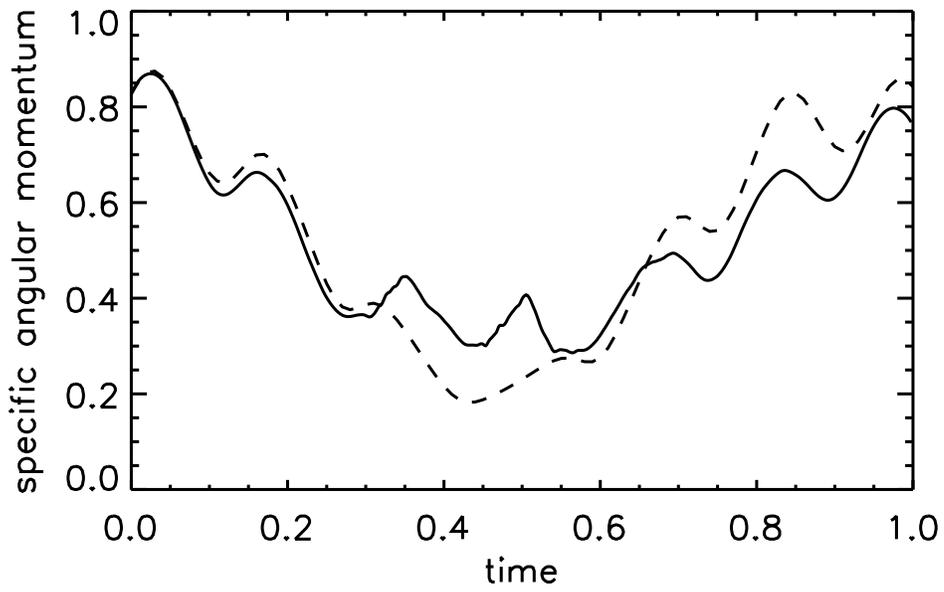
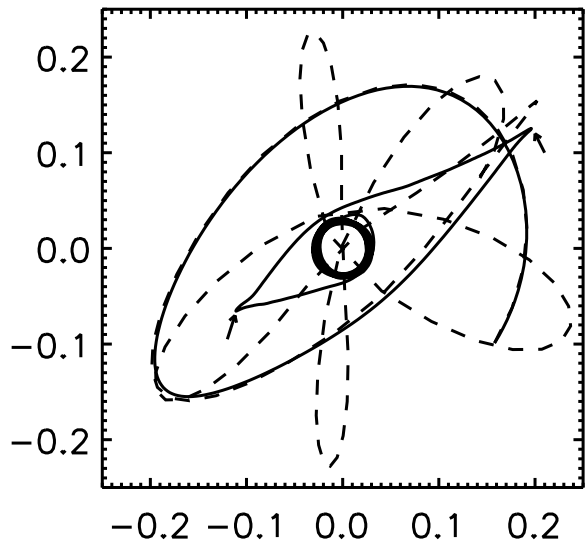
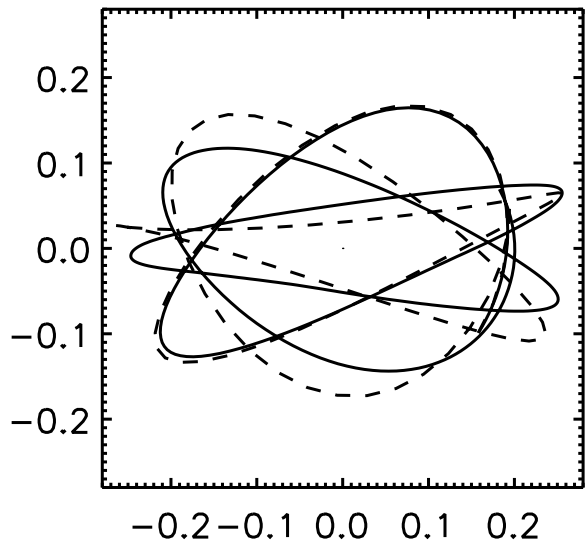
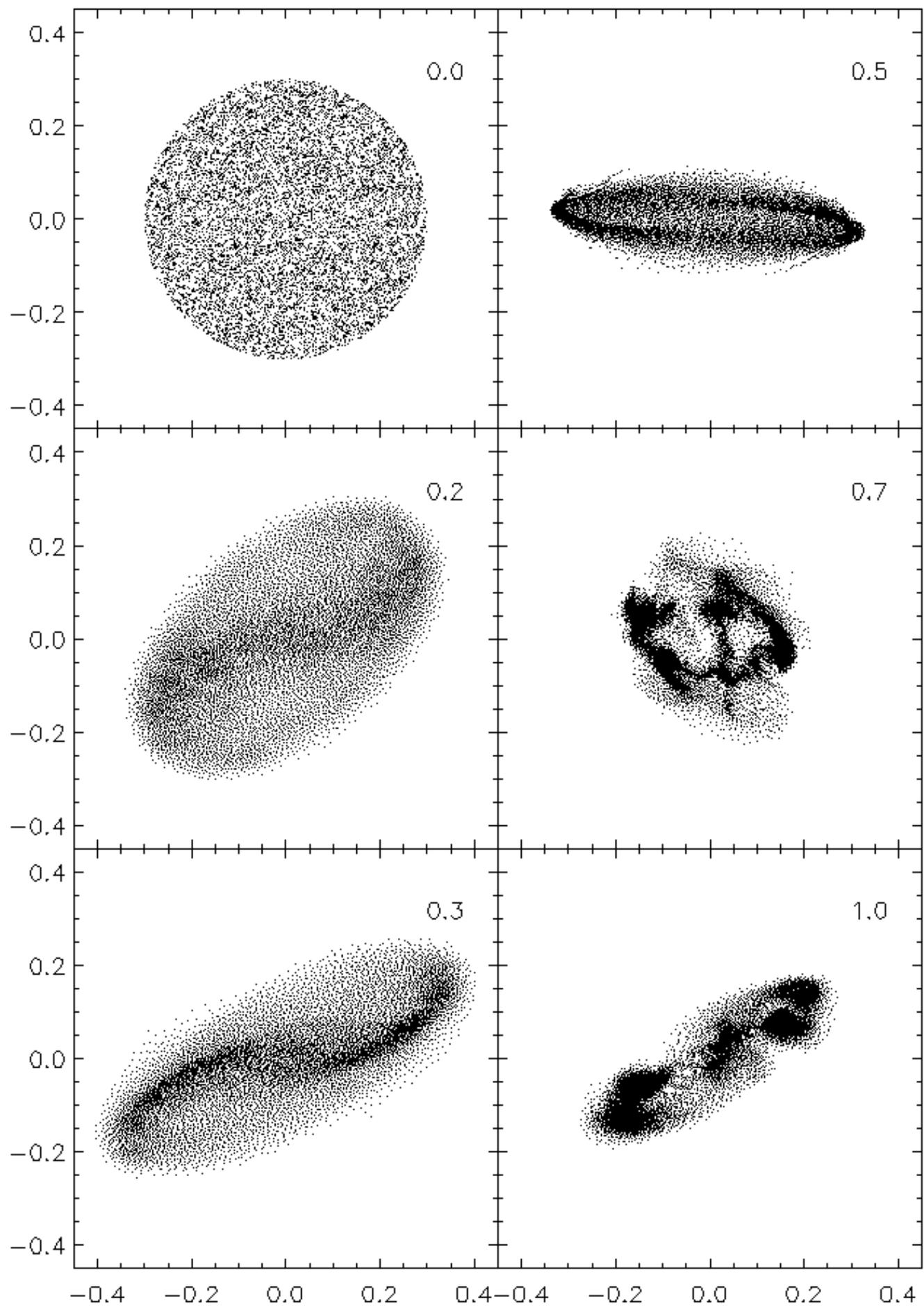
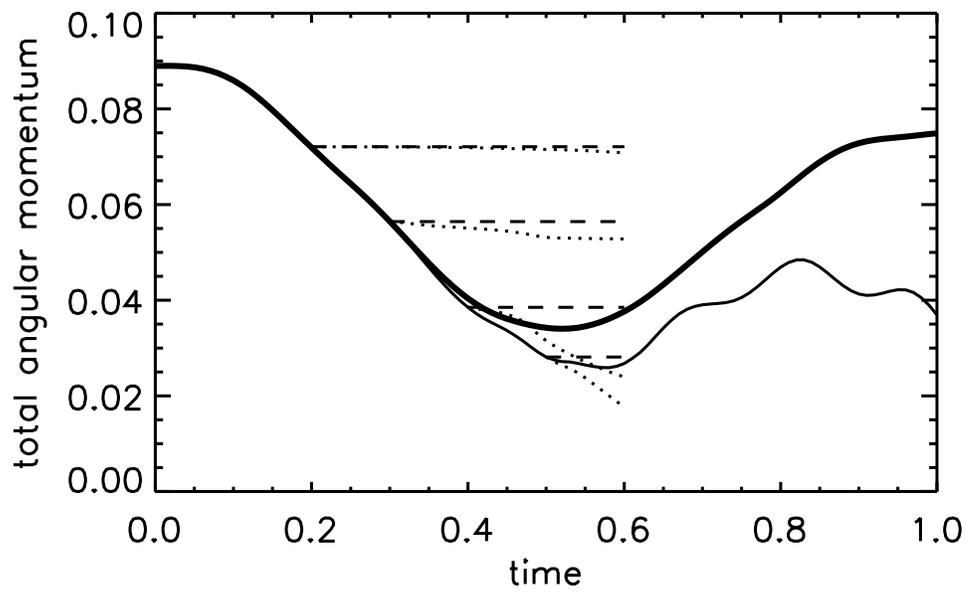
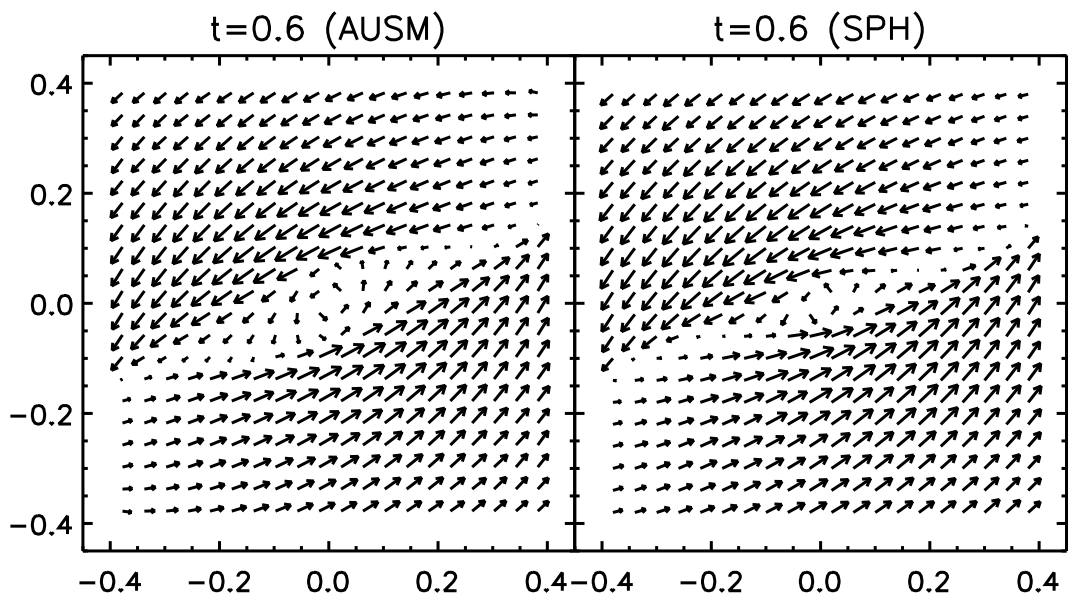
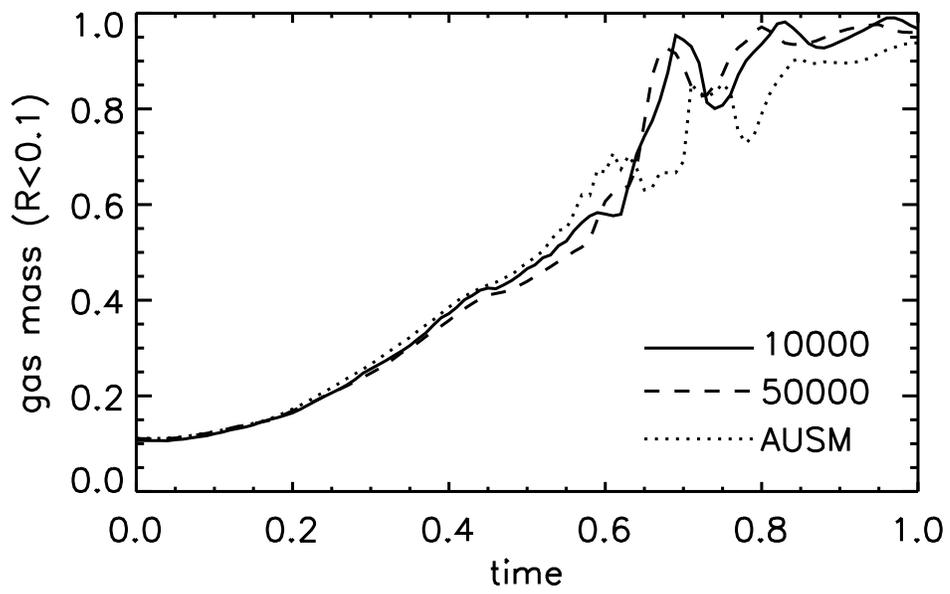
\end{document}